\documentclass[12pt]{article}
\usepackage{graphicx}

\newcommand\pubnumber{XXXX-yyy-zz}
\newcommand\pubdate{January 14, 2011}

\def\napoli{Purdue University\\
West Lafayette, Indiana 47907, USA}
\def\support{\footnote{Work supported by the U.S. Department of Energy Grant DE-FG02-91ER40681A29 to Purdue University.
}}

\def\Title#1{\begin{center} {\Large #1 } \end{center}}
\def\Author#1{\begin{center}{ \sc #1} \end{center}}
\def\Address#1{\begin{center}{ \it #1} \end{center}}

\newcommand\pubblock{\rightline{\begin{tabular}{l} \pubnumber\\
         \pubdate  \end{tabular}}}
\newenvironment{Abstract}{\begin{quotation}  }{\end{quotation}}
\newenvironment{Presented}{\begin{quotation} \begin{center}
             PRESENTED AT\end{center}\bigskip
      \begin{center}\begin{large}}{\end{large}\end{center} \end{quotation}}
\def\Acknowledgements{\bigskip  \bigskip \begin{center} \begin{large}
             \bf ACKNOWLEDGEMENTS \end{large}\end{center}}

\def\beq{\begin{equation}}
\def\eeq#1{\label{#1}\end{equation}}
\def\eeqn{\end{equation}}

\def\beqa{\begin{eqnarray}}
\def\eeqa#1{\label{#1}\end{eqnarray}}
\def\eeqan{\end{eqnarray}}

\let\bar=\overbar

\def\Dslash{\not{\hbox{\kern-4pt $D$}}}
\def\dslash{\not{\hbox{\kern-2pt $\del$}}}

\def\msb{{\bar{\ssstyle M \kern -1pt S}}}

\newcommand{\vcq}{V_{cd(s)}}

\newcommand{\ipb}{{\rm pb}^{-1}}

\newcommand{\bmath}{\begin{displaymath}}
\newcommand{\emath}{\end{displaymath}}

\newcommand{\mbc}{M_{{\rm BC}}}

\newcommand{\EE}{e^+e^-}

\newcommand{\vcs}{|V_{cs}|}
\newcommand{\vcd}{|V_{cd}|}

\newcommand{\beqn}{\begin{equation}}

\newcommand{\fd}{f_{D^+}}
\newcommand{\fds}{f_{D^+_s}}

\newcommand{\dtoenu}{D^+\to e^+\nu}
\newcommand{\dtomunu}{D^+\to \mu^+\nu}
\newcommand{\dtotaunu}{D^+\to \tau^+\nu}

\newcommand{\dstoenu}{D^+_s\to e^+\nu}
\newcommand{\dstomunu}{D^+_s\to \mu^+\nu}
\newcommand{\dstotaunu}{D^+_s\to \tau^+\nu}

\newcommand{\tautopinu}{\tau^+\to \pi^+\bar{\nu}}
\newcommand{\tautoenunu}{\tau^+\to e^+\nu \bar{\nu}}
\newcommand{\tautorhonu}{\tau^+\to \rho^+\nu}
\newcommand{\tautomununu}{\tau^+\to \mu^+\nu \bar{\nu}}

\newcommand{\dstolnu}{D_s^+\to \ell^+\nu}

\usepackage{graphicx}
\usepackage{dcolumn}
\usepackage{amsmath}
\usepackage{epsfig}
\usepackage{relsize}
\RequirePackage{xspace}

\def\cleoc{\hbox{CLEO-c}}
\def\babar{\mbox{\slshape B\kern-0.1em{\smaller A}\kern-0.1em B\kern-0.1em{\smaller A\kern-0.2em R}}}

\begin{document}
\begin{titlepage}
\pubblock

\vfill
\Title{Leptonic Charm Decays}
\vfill
\Author{Bo Xin\support}
\Address{\napoli}
\vfill
\begin{Abstract}
We review the recent experimental results on $D$ and $D_s$ meson leptonic decays from \cleoc, Belle, and \babar,
which results in the decay constants $\fd$= (206.7 $\pm$ 8.9) MeV and $\fds$= (257.3 $\pm$ 5.3) MeV.
The latter is an average obtained by the Heavy Flavor Averaging Group (HFAG).
Comparisons with Lattice QCD (LQCD) calculations are discussed.
\end{Abstract}
\vfill
\begin{Presented}
The 6th International Workshop on the CKM Unitarity Triangle\\
University of Warwick, UK, September 6---10, 2010
\end{Presented}
\vfill
\end{titlepage}
\def\thefootnote{\fnsymbol{footnote}}
\setcounter{footnote}{0}

\section{Introduction}

Leptonic decays are a very clean way to access QCD.
In the Standard Model (SM), the leptonic decays occur
via the annihilation of the constituent quarks of the parent meson
into a virtual $W^+$ boson that subsequently materializes as a lepton-antineutrino pair.
The leptonic decay rate of a $D^+$ or $D_s^+$ meson is given by
\begin{equation}
\Gamma(D_{(s)}^+\to \ell^+\nu) = {{G_F^2}\over
8\pi}f_{D_{(s)}}^2m_{\ell}^2M_{D_{(s)}^+} \left(1-{m_{\ell}^2\over
M_{D_{(s)}^+}^2}\right)^2 \left|V_{cq}\right|^2~~~, \label{eq:equ_rate}
\end{equation}
where $G_F$ is the Fermi coupling constant,
$M_{D_{(s)}^+}$ is the mass of the $D^+$ or $D_s^+$, $m_{\ell}$ is the final state lepton mass,
 $|V_{cq}|$ is the CKM matrix element $\vcs$ or $\vcd$, and $f_{D_{(s)}}$ is the ``decay constant," a
parameter related to the overlap of the heavy and light quark wave-functions at zero spatial separation.
In the charm sector, the CKM matrix elements are tightly constrained by unitarity. Using Eq.~(\ref{eq:equ_rate}),
measurements of leptonic decay rates lead to the decay constants and test of QCD calculations.

This ability to test QCD calculations makes the studies of charm leptonic decays valuable to $B$ mixing.
It is not possible to determine $f_B$ from leptonic $B$ decays,
theoretical calculations of $f_B$ must be used to translate measurements of $B$ mixing to CKM matrix elements.
Testing QCD calculations of the decay constants $\fd$ and $\fds$
helps build more confidence in the theoretical calculations of $f_B$. 
In addition, leptonic charm decays are sensitive to new physics - various new physics
	scenarios can affect the leptonic $D$ branching ratios.

\section{Overview of experimental measurements}

Comparing the decay rates for the various leptonic decay channels of the $D$ and $D_s^+$ mesons, we make the following observations.
First, $D_s$ decays have larger branching ratios than $D$ decays, because of the CKM factor $|\vcq|$.
Second, due to the helicity and phase space factors,
$\Gamma(\dtotaunu):\Gamma(\dtomunu):\Gamma(\dtoenu) = 2.65:1:2.3\times 10^{-5}$,
$\Gamma(\dstotaunu):\Gamma(\dstomunu):\Gamma(\dstoenu) = 9.76:1:2.3\times 10^{-5}$.
In both cases, $\tau^+\nu$ gives the largest rate, but we need to deal with at least two neutrinos.
$\mu^+ \nu$ is the cleanest mode, with only one neutrino.
The electron channel is too small, unless there is new physics.
For $\tau^+ \nu$, the following $\tau^+$ decay modes have been utilized
in the experimental measurements:
${\cal B}(\tautopinu)\approx$ 11\%,
${\cal B}(\tautoenunu)\approx$ 18\%,
${\cal B}(\tautorhonu)\approx$ 25\%, and
${\cal B}(\tautomununu)\approx$ 17\%~\cite{PDG2008}.

Next we give an overview of these recent experimental measurements of the decay constants from \cleoc, Belle, and \babar.
The \cleoc ~and \babar~ results have been updated to their full data samples.

\section{$D^+\to \ell^+\nu$ and $D_s^+\to \ell^+\nu$ at \cleoc}

At \cleoc, the leptonic $D$ decays are studied using 818 $\ipb$ of data taken at the center-of-mass energy of $E_{\rm CM}$=3.770 GeV, where $e^+e^-\to D^+D^-$.
The leptonic $D_s$ decays are studied using 600 $\ipb$ of data taken at $E_{\rm CM}$=4.170 GeV.
The $D_s$ mesons used are from the reactions $e^+e^-\to D_s^{*+}D_s^-$ or $D_s^+D_s^{*-}$.

At 3.770 GeV, the analysis strategy is to reconstruct one of the two $D$ mesons in a hadronic final state,
and search for the leptonic decay in the system recoiling from the tag $D$.
To reconstruct the tag $D$, it is required that the tag candidate have a measured energy consistent with the beam energy,
and a \lq\lq beam constrained mass", $\mbc$, consistent with the mass of a $D$ meson,
where $\mbc = \sqrt{E_{\rm beam}^2/c^4-|\vec{p}_D|^2/c^2}$, $E_{\rm beam}$ is the beam energy, and $\vec{p}_D$ is the measured momentum of the tag candidate.
The clean event environment at charm threshold leads to a high tagging efficiency.

After identifying a $D$ tag, the decays $\dtomunu$ are sought by requiring only one additional oppositely charged track within $|\cos\theta|<$0.9,
where $\theta$ is the angle between the track and the positron beam direction, and there is no additional photons with energy greater than 250 MeV.
The selected events are then separated into two cases: case (i), where the charged track deposits less than 300 MeV of energy in the calorimeter,
and case (ii), where the deposited energy is greater than 300 MeV.
Case (i) contains 98.8\% of the muons, and 55\% of the pions.

\begin{figure}[tbp]
\centering
\includegraphics[width=1.9in,height=2in]{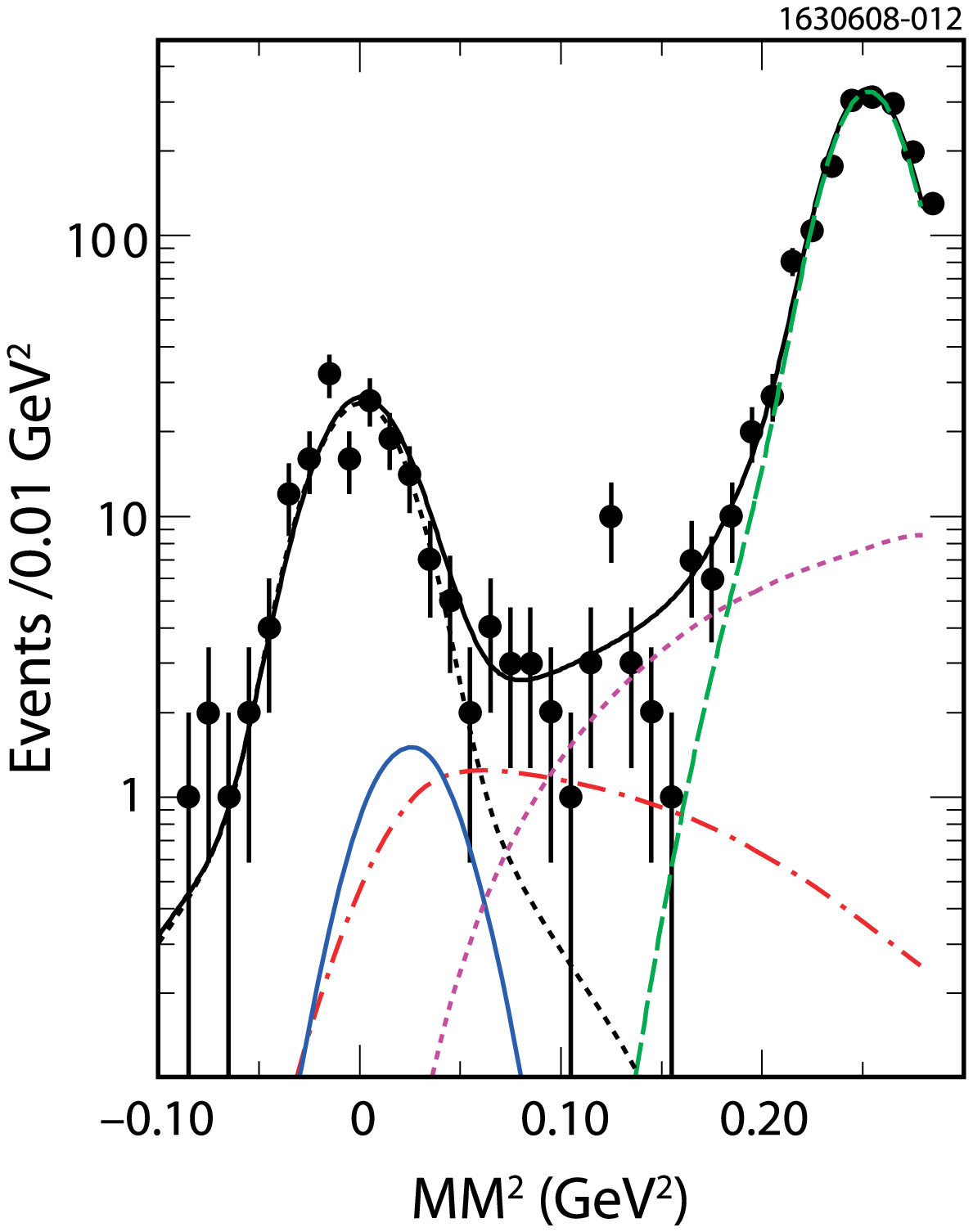}
\includegraphics[width=1.9in,height=1.96in]{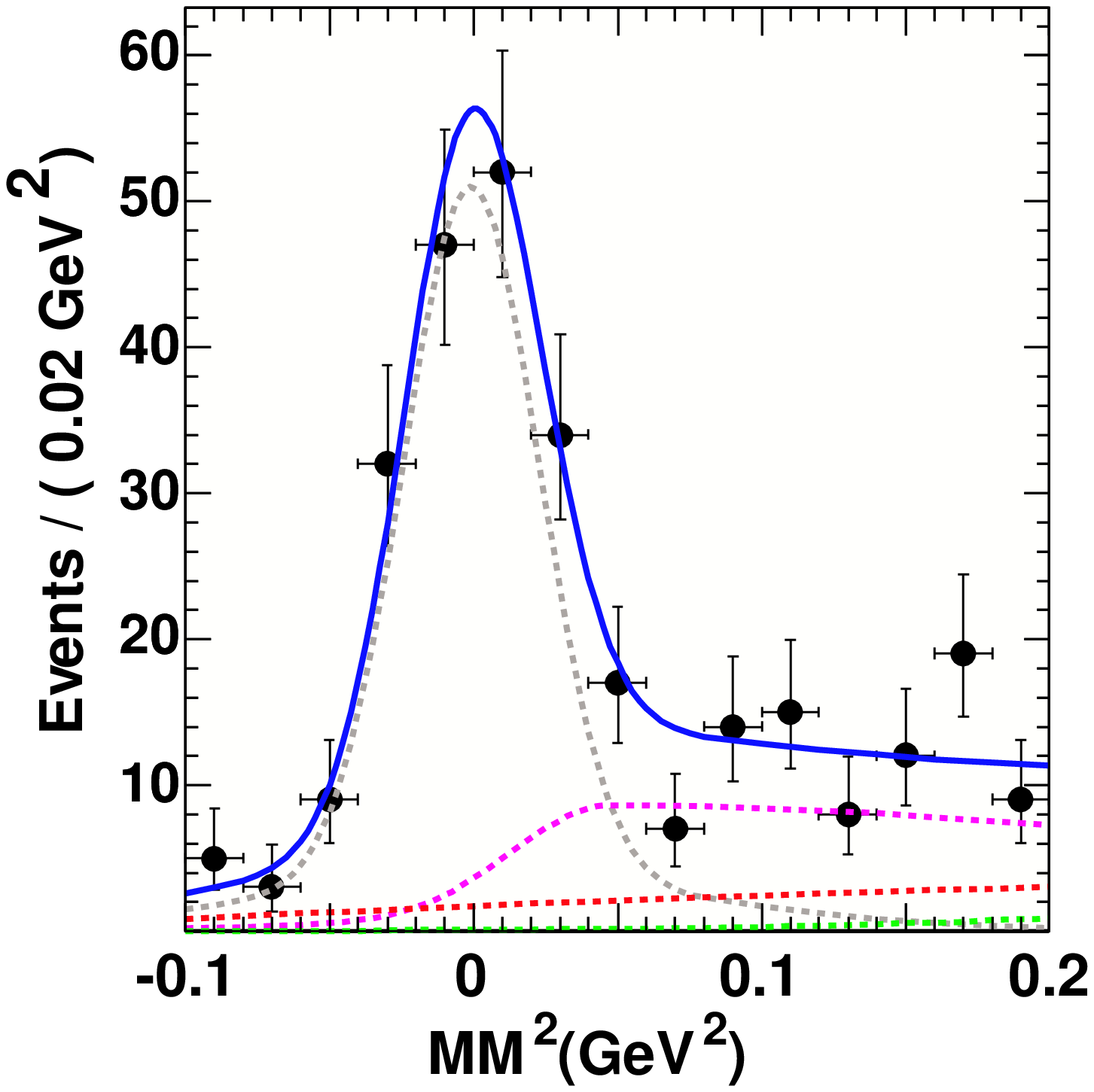}
\includegraphics[width=1.9in,height=2in]{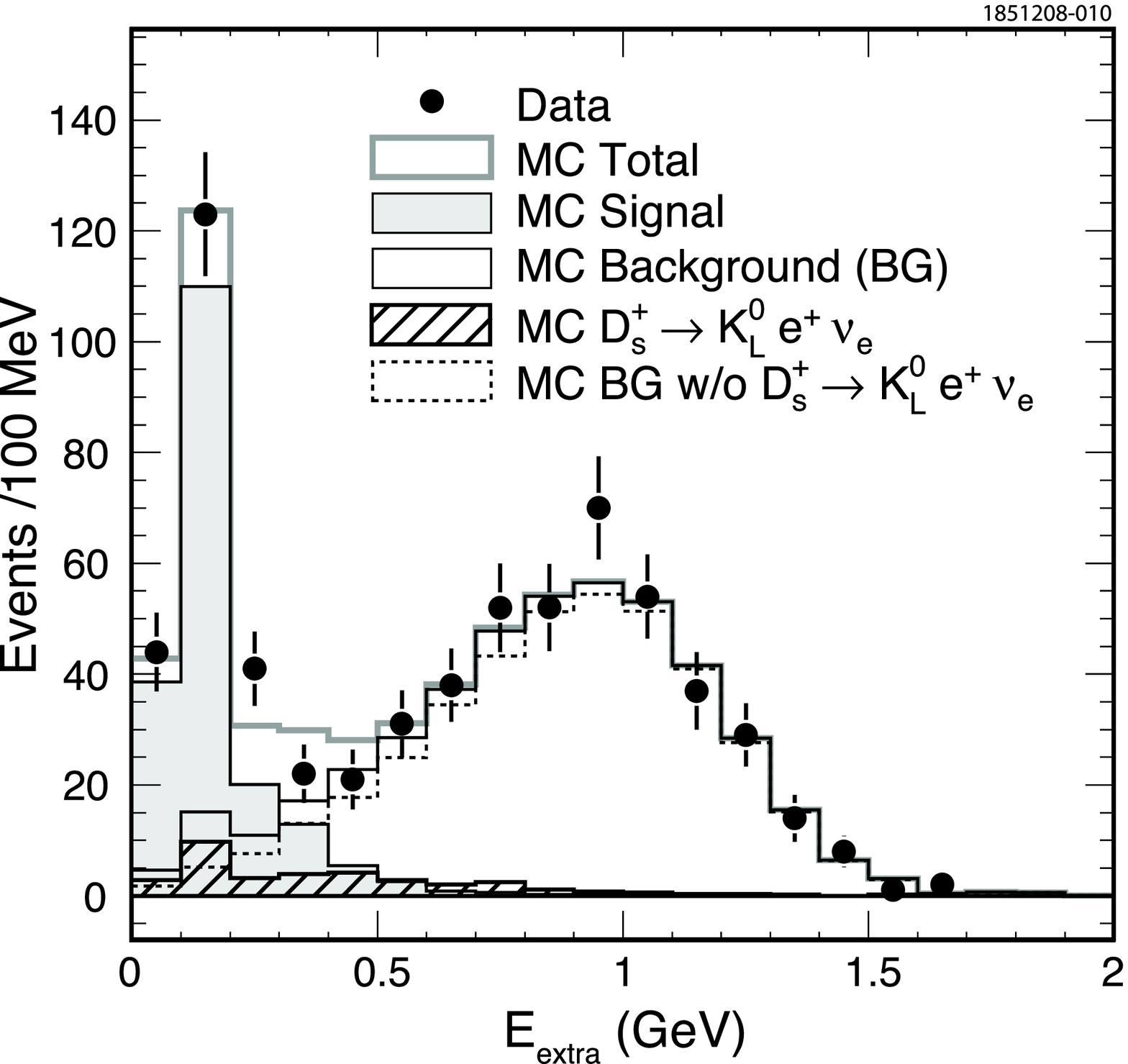}
\caption{Fits to the \cleoc ~missing-mass-squared (MM$^2$) for $\dtomunu$ (left) and $\dstotaunu$, $\tautopinu$ (middle),
and unassociated calorimeter energy ($E_{\rm extra}$) for $\dstotaunu$, $\tautoenunu$ (right).}
\label{fig:munutaunu12}
\end{figure}

The missing-mass-squared is then computed:
\begin{equation}
{\rm MM}^2=\left(E_{\rm beam}-E_{\mu^+}\right)^2-\left(-\vec{
p}_{\rm tag} -\vec{p}_{\mu^+}\right)^2, \label{eq:MMsq}
\end{equation}
where $\vec{p}_{\rm tag}$ is the three-momentum of the fully
reconstructed $D$ tag, and $E_{\mu^+}(\vec{p}_{\mu^+})$ is the energy
(momentum) of the $\mu^+$ candidate.
Fig.~\ref{fig:munutaunu12} (left) shows the MM$^2$ distribution for case (i), which contains separate shapes for signal,
$\pi^+\pi^0$, $\bar{K}^0\pi^+$, $\tau^+\nu$ with $\tau^+\to\pi^+\bar{\nu}$, and a background.
In Fig.~\ref{fig:munutaunu12} (left), the ratio of $\tau^+\nu$, $\tautopinu$ to $\mu^+\nu$ events is fixed to 2.65, as predicted by the SM.
The CLEO collaboration finds
${\cal{B}}(D^+\to\mu^+\nu)=(3.82\pm 0.32\pm 0.09)\times 10^{-4}$ and $f_{D^+}=(206.7\pm 8.5\pm 2.5)$ MeV~\cite{d2munu,munucorrection}.
Upper limits on ${\cal B}(\dtoenu)$ and ${\cal B}(\dtotaunu)$ are also obtained.

$D_s$ tagging at 4.170 GeV is slightly more complicated than $D$ tagging at 3.770 GeV, due to the additional photon
from $D_s^* \to \gamma D_s$.
The reconstruction of the leptonic $D_s$ decays is achieved by first forming a $D_s$ meson from a hadronic final state, then combining it
with a well-reconstructed photon, and calculating the missing-mass-squared (MM$^{*2}$) recoiling against the $D_s + \gamma$ pair. 

On the signal side recoiling against the $D_s+\gamma$ combination,
it is required that one, and only one charged track exists, with charge opposite to the tag $D_s$.
The missing-mass-squared recoiling against the $D_s\gamma$ + track system is then
\begin{equation}
{\rm MM}^{2} = (E_{\rm
CM}-E_{D_s}-E_{\gamma}-E_{\mu})^2-(\vec{p}_{\rm
CM}-\vec{p}_{D_s}-\vec{p}_{\gamma}-\vec{p}_{\mu})^2.
\end{equation}
The events are then separated into three cases:
case (i), where the track deposits less than 300 MeV energy in the calorimeter, characteristic of a muon or a non-interacting pion;
case (ii), where the track deposits more than 300 MeV energy in the calorimeter and does not pass the electron selection, characteristic of an interacting pion;
and case (iii), where the track satisfies the electron selection criteria.
case (i) has 98.8\% of the $\mu^+ \nu$ events and 55\% of the $\tau^+ \nu$, $\tautopinu$ events.
The fit to the case (i) and (ii) combined MM$^2$ distribution is shown in Fig.~\ref{fig:munutaunu12} (middle),
where the $\tau^+\nu/\mu^+\nu$ ratio has been fixed to the SM prediction.
When the $\tau^+\nu/\mu^+\nu$ ratio is allowed to float, the branching fractions are determined to be
${\cal B}(\dstomunu) = (0.565\pm 0.045 \pm 0.017)$\% and
${\cal B}(\dstotaunu)= (6.42 \pm 0.81 \pm 0.18)$\%.
Fixing the $\tau^+\nu/\mu^+\nu$ ratio gives $\fds=(259.5\pm6.6\pm3.1)$ MeV.

The second CLEO analysis on $\dstotaunu$ utilizes $\tautoenunu$. 
The technique is to search for events with one electron and not much other energy opposite a $D_s$ tag.
To ensure an extremely clean event environment, only the three cleanest tag modes are used.
It is required that $E_{\rm extra}$, the sum of the extra energy in the calorimeter not associated with the $D_s$ tag or the electron,
to be less than 400 MeV.
Note that there is no need to find the photon from the $D_s^*$.
Fig.~\ref{fig:munutaunu12} (right) shows the distribution of $E_{\rm extra}$. 
Most of the backgrounds is due to semileptonic decays, but these are outside of the signal region.
The CLEO collaboration finds
${\cal B}(\dtotaunu)=(5.30\pm0.47\pm0.22)$\% and $\fds=(252.5\pm11.1\pm5.2)$ MeV~\cite{ds2taunuenunu}.

The third CLEO analysis on $\dstotaunu$ uses $\tautorhonu$~\cite{ds2taunurhonu}.
The tagging technique used here is identical to the $\dstotaunu$, $\tautopinu$ analysis.
However,  
because there are two neutrinos, the MM$^2$ does not peak for the signal, but does peak for the important backgrounds.
The branching fractions of the peaking backgrounds are pre-measured using a double tag technique with the same set of $D_s$ tags.
The sum of the extra energy in the calorimeter ($E_{\rm extra}$) is used an important discriminant.
Signal yields are extracted by simultaneously fitting to the $D_s$ tag invariant mass and the
MM$^2$ distributions, separately in three $E_{\rm extra}$ intervals:
(i) $E_{\rm extra}>$ 800 MeV, where signal is absent,
(ii) $E_{\rm extra}<$ 100 MeV, where signal dominates,
and (iii) $E_{\rm extra} \in$  (100, 200) MeV, where signal and background are about equal.
The interval $E_{\rm extra}>$ 800 MeV is used as the background control sample. 
The branching fraction of $\dstotaunu$ is measured in two $E_{\rm extra}$ bins where signal is present.
The results are then combined, resulting in ${\cal B}(\dstotaunu)=(5.52\pm0.57\pm0.21)$\%
and $\fds=(257.8\pm13.3\pm5.2)$\%.

\section{$D_s^+\to \ell^+\nu$ at the $B$ factories}

\begin{figure}[tb]
\centering
\includegraphics[width=2.5in,height=2in]{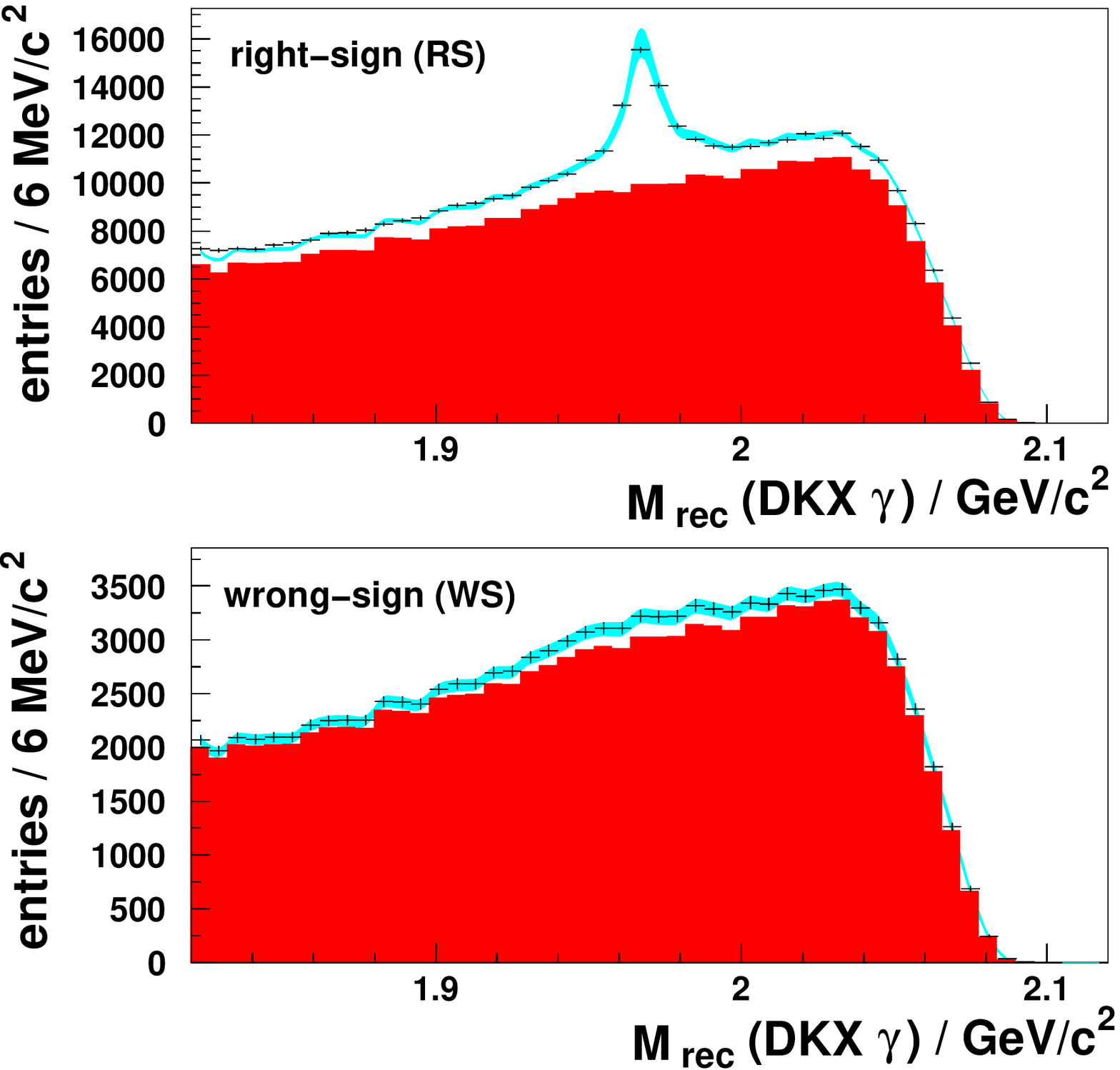}
\includegraphics[width=2.5in,height=2in]{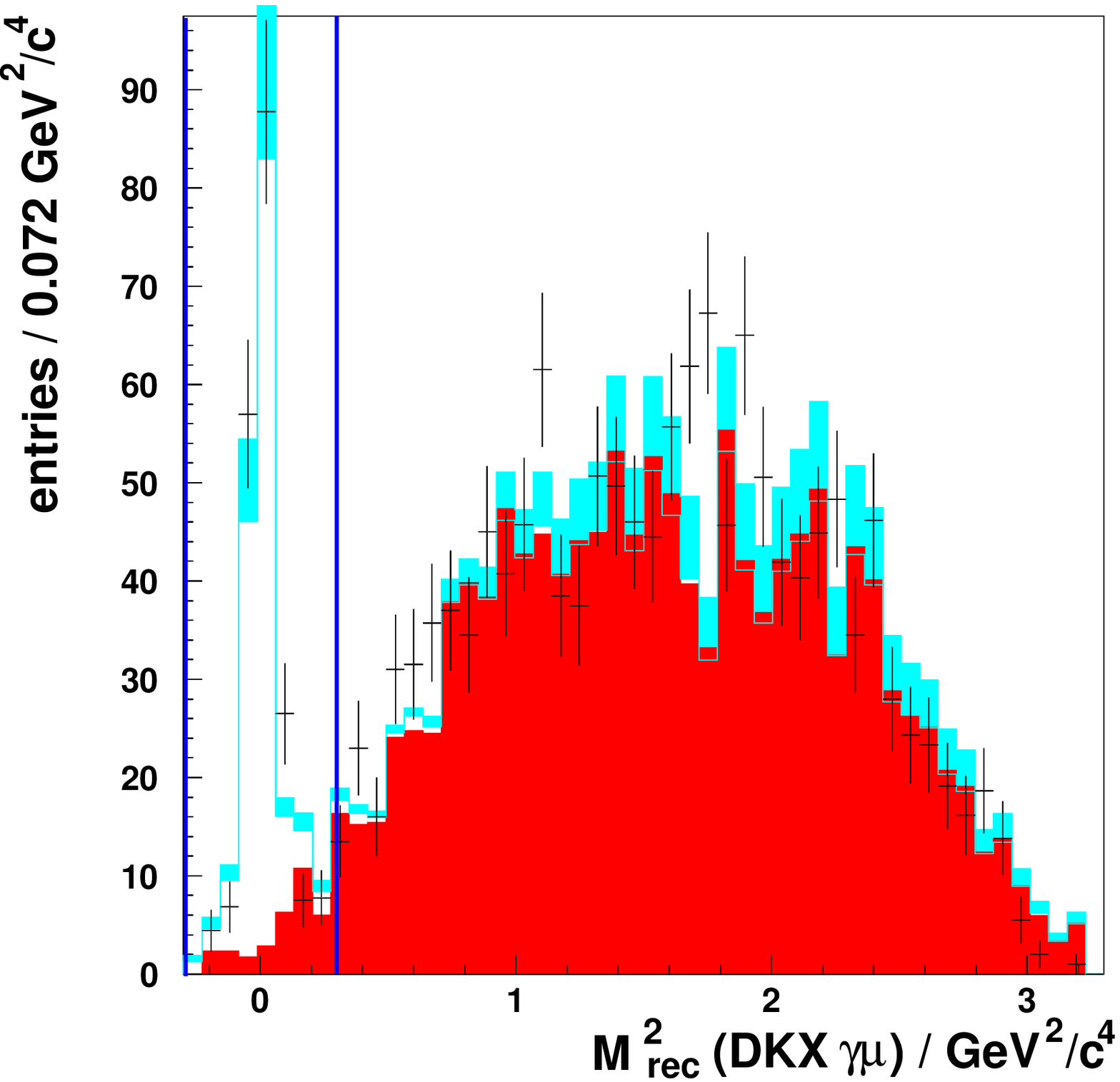}
\caption{Recoil mass for $D_s$ tags (left) and
missing-mass-squared for $D_s^+\to\mu^+\nu_\mu$ candidates (right)
from the Belle $\dstomunu$ analysis~\cite{bellemunu}.
}
\label{fig:belle}
\end{figure}

To measure the branching fraction of $\dstomunu$ and $\fds$,
Belle has performed a full event reconstruction using their 548 fb$^{-1}$ data sample~\cite{bellemunu}.
They seek events of the type $\EE\to D_s^* D^{\pm,0} K^{\pm,0}X$, where $D_s^*\to \gamma D_s$,
and $X$ can be any number of additional pions from fragmentation, and up to one photon.
The four-momentum of the $D_s$ is determined without ambiguity by reconstruction of the system recoiling against the $DK\gamma X$ in the event.
Fig.~\ref{fig:belle} (left) shows the distributions of recoil mass against the $DK\gamma X$. 

After further identifying a muon candidate, the recoil-mass-squared against the $DK\gamma X\mu$ is calculated. 
Fig.~\ref{fig:belle} (right) shows the missing-mass-squared distribution.
This results in the branching fraction ${\cal B}(\dstomunu)=(6.44\pm0.76\pm0.57)$\%,
implying the decay constant $\fds=(275\pm16\pm12)$ MeV~\cite{bellemunu}.

A recent BABAR analysis on $\dstolnu$~\cite{babarlnu} uses a technique similar to the one Belle uses in their measurement of $\dstomunu$~\cite{bellemunu}.
An inclusive sample of $D_s$ is obtained by reconstructing the rest of the event in reactions of the kind $\EE\to DKXD_s^*$, where $D_s^*\to D_s\gamma$.
The $\dstomunu$ signal is then reconstructed by computing the recoil-mass-squared against the $DKX\gamma\mu$. 
For $\dstotaunu$, BABAR used $\tautoenunu$ and $\mu^+\nu\bar{\nu}$ channels.
The extra energy in the calorimeter ($E_{\rm extra}$) is used to reconstruct these $\tau^+ \nu$ events.
The BABAR collaboration finds ${\cal B}(\dstomunu)=(6.02\pm0.38\pm0.34)$\%
and ${\cal B}(\dstotaunu)=(5.00\pm0.35\pm0.49)$\%, averaged over $\tautoenunu$ and $\mu^+\nu\bar{\nu}$.
Using these results, the decay constant is measured to be $\fds=(258.6\pm6.4\pm7.5)$ MeV~\cite{babarlnu}.

\section{Conclusions}

\begin{figure}[tb]
\centering
\includegraphics[width=3.5in]{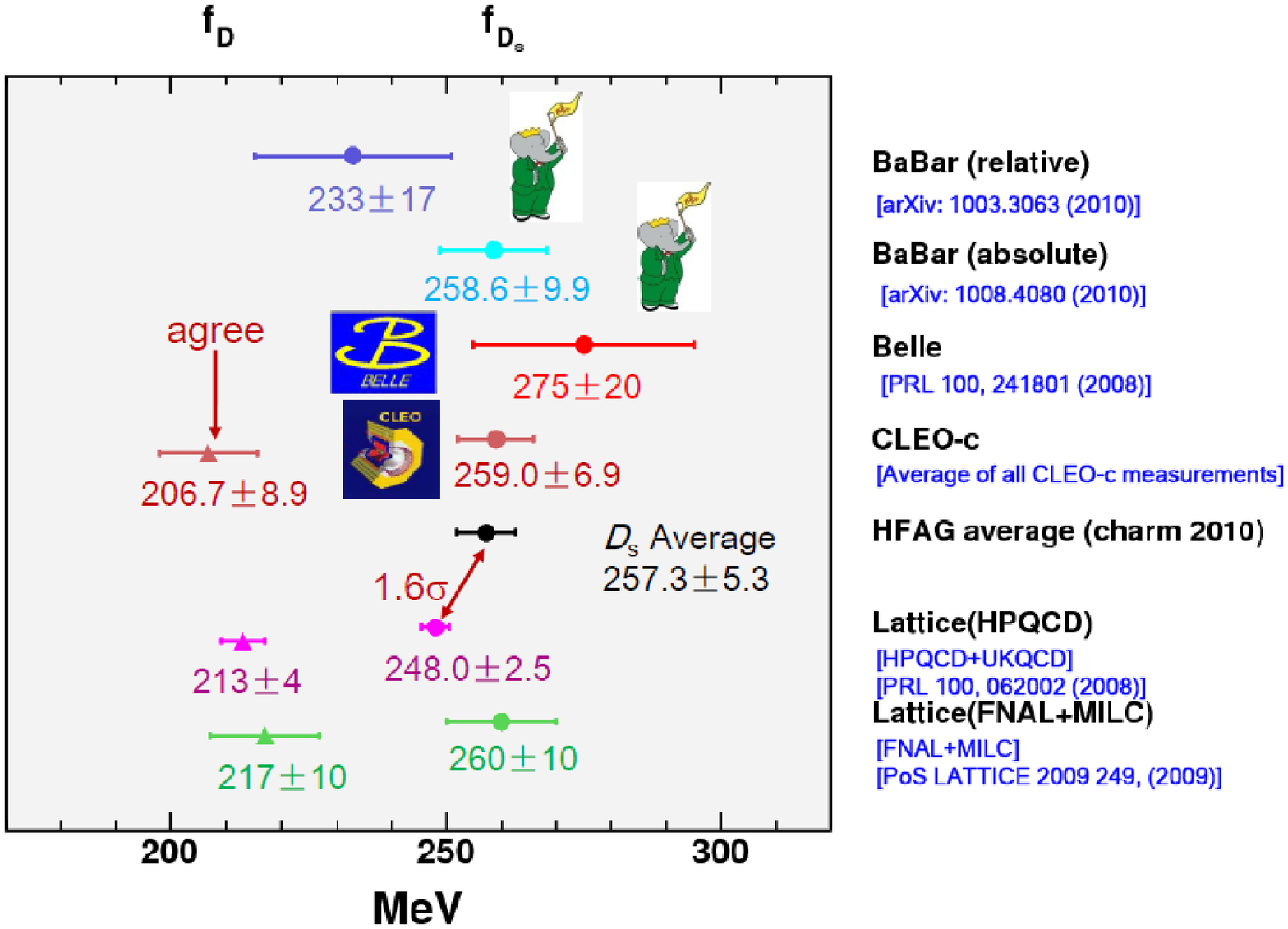}
\caption{Comparison of the $\fds$ results between experiments and LQCD calculations.}
\label{fig:sum}
\end{figure}

Precision measurements of charm decay constants at \cleoc, Belle, and \babar~ have brought the test of LQCD calculations to an unprecedented level.
As a result, the past a few years witnessed considerable tension between theory and experiment with regard to the decay constant $\fds$.
With the dramatic improvements in precisions of both experimental measurements and LQCD calculations,
we have seen disagreement between LQCD and experiment as large as 3.8$\sigma$~\cite{Kronfeldpuzzle}.
This led to much speculation about the existence of new physics.
Fig.~\ref{fig:sum} gives a summary of the current status.

An average of all the \cleoc ~measurements gives $\fds=(259.0\pm6.2\pm3.0)$ MeV~\cite{ds2taunurhonu}.
The \babar ~results in Ref.~\cite{babarlnu} are correlated with another \babar~ analysis~\cite{babartaunu}, 
which we haven't described here.
The Heavy Flavor Averaging Group (HFAG) has averaged all the recent experimental results on $\fds$ and obtains
$\fds=(257.3\pm5.3)$ MeV~\cite{hfagfds}.
Results from $\dstomunu$ and $\dstotaunu$ are consistent.
A recent HPQCD calculation gives 
$\fds=(248.0\pm2.5)$ MeV~\cite{hpqcd1008},
which 
differ from the HFAG average by 1.6$\sigma$.
Another calculation by the Fermilab Lattice and MILC Collaborations is also compatible with experiments but has larger uncertainties~\cite{fnalmilcfds}.

Theory and experiment agree well on $\fd$.
Experiments have achieved 4.3\% precision on $\fd$ and 2.1\% on $\fds$.
Should the difference between theory and experiment on $\fd$ or $\fds$ be found to be significant in the future, it could be a signal of physics beyond the SM.
With the planned $\sim$10 fb$^{-1}$ open charm data at BESIII, more stringent tests of LQCD are expected.

\Acknowledgements

I thank the organizers for such a wonderful conference at the University of Warwick.
Valuable discussions with L. M. Zhang, H. Na, and Z. Liu are appreciated.
I. Shipsey is thanked for very helpful discussions and suggestions on this manuscript.

\end{document}